\documentclass[pre,preprint,endfloats,amsmath,superscriptaddress]{revtex4}

\begin{document}

\title{The Onset of Synchronization in Systems of Globally Coupled Chaotic and Periodic Oscillators}

\author{Edward Ott}
\affiliation{Institute for Research in Electronics and Applied Physics, Department of Physics, and Department of Electrical and Computer Engineering,
University of Maryland, College Park, Maryland, 20742}

\author{Paul So}
\affiliation{Department of Physics and Astronomy and the Krasnow Institute for Advanced Study, George Mason University, Fairfax, Virginia, 22030}

\author{Ernest Barreto}
\affiliation{Department of Physics and Astronomy and the Krasnow Institute for Advanced Study, George Mason University, Fairfax, Virginia, 22030}

\author{Thomas Antonsen}
\affiliation{Institute for Research in Electronics and Applied Physics, Department of Physics, and Department of Electrical and Computer Engineering,
University of Maryland, College Park, Maryland, 20742}


\begin{abstract}
A general stability analysis is presented for the determination of the transition from incoherent to coherent behavior in an ensemble of globally coupled, heterogeneous, continuous-time dynamical systems. The formalism allows for the simultaneous presence of ensemble members exhibiting chaotic and periodic behavior, and, in a special case, yields the Kuramoto model for globally coupled periodic oscillators described by a phase. Numerical experiments using different types of ensembles of Lorenz equations with a distribution of parameters are presented.  
\end{abstract}

\date{\today}


\maketitle

\section{\label{Introduction}Introduction}

Systems of many coupled dynamical units are of great interest in a wide variety of scientific fields including physics, chemistry and biology.  In this paper, we are interested in the case of {\em global coupling} in which each element is coupled to all others. Beginning with the work of Kuramoto \cite{kur84} and Winfree \cite{win80}, there has been much research on synchrony in systems of globally coupled limit cycle oscillators (e.g., \cite{str92,erm91,ace00,yeu99,koz00,red98,fra01,str00}).  Possible applications include groups of chirping crickets\cite{wal69}, flashing fireflies \cite{buc88}, Josephson junction arrays \cite{wie96}, semiconductor laser arrays \cite{red98}, and cardiac pacemaker cells \cite{pes75}.  Recently, Pikovsky, et al.~\cite{pik96} and Sakaguchi \cite{sak00} have studied the onset of synchronization in systems of globally coupled {\em chaotic} systems.

In this paper we present and apply a formal analysis of the stability of the unsynchronized state (or ``incoherent state'') of a general system of globally coupled heterogeneous, continuous-time dynamical systems.  In our treatment, no {\it a priori} assumption about the dynamics of the individual coupled elements is made.  Thus the systems can consist of elements whose natural uncoupled dynamics is chaotic or periodic, including the case where both types of elements are present.  Our treatment is related to the marginal stability investigation of Ref.~\cite{top01}; see also \cite{sak00}.  The main difference between our work and these previous works is that we treat an ensemble of nonidentical systems, considering both chaotic and limit cycle dynamics, and that our work yields growth rates as well as instability conditions.  In addition, our treatment addresses some basic issues of the linear theory (e.g., analytic continuation of the dispersion function).

The organization of the rest of this paper is as follows.  The problem is formulated in Sec.~\ref{Formulation}, and a formal solution for the dispersion relation $D(s)=0$ is given in Sec.~\ref{StabilityAnalysis}.  Here the quantity $s$ governs the stability of the system ($\mbox{Re}(s)>0$ implies instability). The interpretation, analytic properties, and numerical calculation of the dispersion relation are discussed in Sec.~\ref{Discussion} along with other issues related to the treatment given in Sec.~\ref{StabilityAnalysis}.  In Sec.~\ref{Kuramoto}, we obtain $D(s)$ for the Kuramoto model of coupled limit cycle oscillators as an example.  Section \ref{NumericalExperiments} presents illustrative numerical examples using three different ensembles of globally coupled Lorenz equations.  In particular, these ensembles are formed of systems with a parameter $r$ that is uniformly distributed in an interval $[r_-,r_+]$ with three different choices of $r_\pm$.  In the first example (Sec.~\ref{NumExpts:periodic}) every element of the uncoupled ensemble is periodic, but the interval $[r_-,r_+]$
includes a pitchfork bifurcation.  The second example (Sec.~\ref{NumExpts:chaotic}) is for an apparently chaotic ensemble, while the third example (Sec.~\ref{NumExpts:mixed}) involves an ensemble that includes both chaotic and periodic elements.  Finally, Sec.~\ref{Further} concludes the paper with further discussion and a summary of the results.  

\section{\label{Formulation}Formulation}

We first treat the simplest case, giving generalizations later in the paper (Sec.~\ref{Discussion:generalizations}).  We consider dynamical systems of the form
\begin{equation}
d{\bf x}_i(t)/dt={\bf G}({\bf x}_i(t),{\bf \Omega }_i)+{\bf K}(\langle \langle {\bf x}\rangle \rangle _*-\langle \langle
{\bf x}(t)\rangle \rangle ),
\label{first}
\end{equation}
where ${\bf x}_i=(x_i^{(1)},x_i^{(2)},\ldots, x_i^{(q)})^T$ is a $q$-dimensional vector; ${\bf G}$ is a
$q$-dimensional vector function; ${\bf K}$ is a constant $q\times q$ coupling matrix; $i=1,2,\cdots ,N$ is an index labeling components in the ensemble of coupled systems (in our analytical work we take the limit $N \rightarrow \infty$, while in our numerical work $N >> 1$ is finite); $\langle \langle {\bf x}(t)\rangle \rangle $ is the instantaneous average component state (also referred to as the {\em order parameter}),
\begin{equation}
\langle \langle {\bf x}(t)\rangle \rangle =\lim _{N\rightarrow \infty} N^{-1}\sum _i \langle {\bf x}_i(t)\rangle,
\label{order}
\end{equation}
and, for each $i$, $\langle {\bf x}_i\rangle$ is the average of ${\bf x}_i$ over an infinite number of initial conditions ${\bf x}_i(0)$, distributed according to some chosen initial distribution on the attractor of the $i$th uncoupled system
\begin{equation}
d{\bf x}_i/dt={\bf G}({\bf x}_i,{\bf \Omega }_i).
\label{ithuncoupled}
\end{equation}
${\bf \Omega }_i$ is a parameter vector specifying the uncoupled $({\bf K}=0)$ dynamics, and $\langle \langle {\bf
x}\rangle \rangle _*$ is the {\it natural measure} \cite{ott93} and $i$ average of the state of the uncoupled system.  That is, to
compute $\langle \langle {\bf x}\rangle \rangle_*$, we set ${\bf K}=0$, compute the solutions to Eq.~(\ref{ithuncoupled}), and
obtain $\langle \langle {\bf x}\rangle \rangle_*$ from 
\begin{subequations}
\label{xstar}
\begin{equation}
\langle \langle {\bf x}\rangle \rangle _*=\lim _{N\rightarrow \infty}N^{-1}\sum _i [\lim _{\tau _0\rightarrow
\infty}\tau _0^{-1}\int ^{\tau _{0}}_0{\bf x}_i(t)dt].
\end{equation}
In what follows we assume that the ${\bf
\Omega}_i$ are randomly chosen from a smooth probability density function $\rho ({\bf \Omega })$. Thus, an alternate means of expressing (\ref{xstar}a) is
\begin{equation}
\langle \langle {\bf x}\rangle \rangle _*=\int {\bf x}\rho({\bf \Omega}) d\mu_{\bf \Omega} d{\bf \Omega},
\end{equation}
\end{subequations}
where $\mu_{\bf \Omega}$ is the natural invariant measure for the system $d{\bf x}/dt = {\bf G}({\bf x}, {\bf \Omega})$. By
construction, $\langle \langle {\bf x}\rangle \rangle =\langle \langle {\bf x}\rangle \rangle _*$ is a solution of
the globally coupled system (\ref{first}).  We call this solution the ``incoherent state'' because the coupling term cancels and the individual oscillators do not affect each other.  The question we address is whether the incoherent state is stable.  In particular, as a system parameter such as the coupling strength varies, the onset of instability of the incoherent state signals the start of coherent, synchronous behavior of the ensemble.  

\section{\label{StabilityAnalysis}Stability Analysis}

To perform the stability analysis, we assume that the system is in the incoherent state, so that at any fixed time $t$, and for each $i$, ${\bf x}_i(t)$ is distributed according to the natural measure. We then perturb the orbits ${\bf x}_i(t)\rightarrow {\bf x}_i(t)+\delta {\bf x}_i(t)$, where $\delta {\bf x}_i(t)$ is an infinitesimal perturbation:
\begin{equation}
d\delta {\bf x}_i/dt={\bf DG}({\bf x}_i(t),{\bf \Omega }_i)\delta {\bf x}_i-{\bf K}\langle \langle \delta {\bf
x}_i\rangle \rangle
\label{perturb}
\end{equation}
where
\[
{\bf DG}({\bf x}_i(t),{\bf \Omega }_i) \delta {\bf x}_i = \delta {\bf x}_i \cdot \frac{\partial}{\partial {\bf x}_i} {\bf G}({\bf x}_i(t),{\bf \Omega }_i).
\]
Introducing the fundamental matrix ${\bf M}_i(t)$ for system (\ref{perturb}),
\begin{equation}
d{\bf M}_i/dt={\bf DG \cdot M}_i,
\label{variationaleq}
\end{equation}
where ${\bf M}_i(0)\equiv \openone$, we can write the solution of Eq.~(\ref{perturb}) as
\begin{equation}
\delta {\bf x}_i(t)=-\int ^t_{-\infty}{\bf M}_i(t){\bf M}^{-1}_i(\tau ){\bf K}\langle \langle \delta {\bf
x}\rangle \rangle _\tau d\tau,
\label{intmm}
\end{equation}
where we use the notation $\langle \langle \delta {\bf x}\rangle \rangle _\tau $ to signify that $\langle \langle \delta {\bf x}\rangle \rangle $ is evaluated at time $\tau $.  Note that, through Eq.~(\ref{variationaleq}), ${\bf M}_i$ depends on the unperturbed orbits ${\bf x}_i(t)$ of the uncoupled nonlinear system (\ref{ithuncoupled}), which are determined by their initial conditions ${\bf x}_i(0)$ (distributed according to the natural measure).  

Assuming that the perturbed order parameter evolves exponentially in time (i.e., $\langle \langle \delta {\bf x}\rangle \rangle ={\bf \Delta }e^{st}$), Eq.~(\ref{intmm}) yields
\begin{equation}
\{ \openone +\tilde {\bf M}(s){\bf K}\}{\bf \Delta }=0,
\label{dispeq_nodet}
\end{equation}
where $s$ is complex, and
\begin{equation}
\tilde {\bf M}(s)=\left\langle \left\langle \int ^t_{-\infty}e^{-s(t-\tau )}{\bf M}_i(t){\bf M}_i^{-1}(\tau )d\tau
\right\rangle\right\rangle_*.
\label{mtilde}
\end{equation}
Thus the dispersion function determining $s$ is 
\begin{equation}
D(s)=\det \{\openone +\tilde {\bf M}(s){\bf K}\}=0.
\label{dispersioneq}
\end{equation}

In order for Eqs.~(\ref{dispeq_nodet}) and (\ref{dispersioneq}) to make sense, the right side of Eq.~(\ref{mtilde}) must be independent of time.  As written, it may not be clear that this is so.  We now demonstrate this, and express $\tilde {\bf M}(s)$ in a more convenient form.  To do this, we make the dependence of ${\bf M}_i$ in Eq.~(\ref{mtilde}) on the initial condition explicit: ${\bf M}_i(t){\bf M}_i^{-1}(\tau )={\bf M}_i(t,{\bf x}_i(0)){\bf M}_i^{-1}(\tau ,{\bf
x}_i(0))$.  From the definition of ${\bf M}_i$, we have
\begin{equation}
{\bf M}_i(t,{\bf x}_i(0)){\bf M}_i^{-1}(\tau ,{\bf x}_i(0))={\bf M}_i(t-\tau ,{\bf x}_i(\tau ))={\bf M}_i(T,{\bf
x}_i(t-T)),
\label{mminv}
\end{equation}
where we have introduced $T=t-\tau $.  Using Eq.~(\ref{mminv}) in Eq.~(\ref{mtilde}) we have
\[
\tilde {\bf M}(s)=\left\langle \left\langle \int ^\infty _0e^{-sT}{\bf M}_i(T,{\bf x}_i(t-T)dT \right\rangle \right\rangle_*.
\]
Note that our solution requires that the integral in the above converge.  Since the growth of ${\bf M}_i$ with increasing $T$ is dominated by $h_i$, the largest Lyapunov exponent for the orbit ${\bf x}_i$, we require
\[
Re(s)>\Gamma \ , \ \ \ \Gamma =\max _{{\bf x}_i,{\bf \Omega }_i}h_i.
\]
In contrast with the chaotic case where $\Gamma >0$, an ensemble of periodic attractors has $\Gamma =0$ (for an attracting periodic orbit $h_i=0$ corresponds to orbit perturbations along the flow).  With the condition $\mbox{Re}(s)>\Gamma $, the integral converges exponentially and uniformly in the quantities over which we
average.  Thus we can interchange the integration and the average,
\begin{equation}
\tilde {\bf M}(s)=\int ^\infty _0e^{-sT}\langle \langle {\bf M}_i(T,{\bf x}_i(t-T))\rangle \rangle_* dT.
\label{mtilde_avginside}
\end{equation}
In Eq.~(\ref{mtilde_avginside}) the only dependence on $t$ is through the initial condition ${\bf x}_i(t-T)$.  However, since the quantity within angle brackets includes not only an average over $i$, but also an average over initial conditions with respect to the natural measure of each uncoupled attractor $i$, the time invariance of the natural measure ensures
that Eq.~(\ref{mtilde_avginside}) is independent of $t$.  In particular, invariance of a measure means that if an infinite cloud of initial conditions ${\bf x}_i(0)$ is distributed on uncoupled attractor $i$ at $t=0$ according to its natural invariant measure, then the distribution of the orbits, as they evolve to any time $t$ via the uncoupled dynamics (Eq.~(\ref{ithuncoupled})), continues to give the same distribution as at time $t=0$.  Hence, although ${\bf M}_i(T,{\bf x}_i(t-T))$ depends on $t$, when we average over initial conditions, the result $\langle {\bf M}_i(T,{\bf x}_i(t-T))\rangle_*$ is independent of $t$ for each $i$.  Thus we drop the dependence of $\langle \langle {\bf M}_i\rangle \rangle_*$ on the initial values of the ${\bf x}_i$ and write
\begin{equation}
\tilde {\bf M}(s)=\int ^\infty _0 e^{-sT}\langle \langle {\bf M}(T)\rangle \rangle_* dT,
\label{mtilde_timeindep}
\end{equation}
where, for convenience we have also dropped the subscript $i$.  Thus $\tilde {\bf M}$ is the Laplace transform of $\langle \langle {\bf M}\rangle \rangle_*$.  This result for $\tilde {\bf M}(s)$ can be analytically continued into $\mbox{Re}(s)<\Gamma$, as explained in Sec.~\ref{Discussion:analytic}.  

Note that $\tilde {\bf M}(s)$ depends only on the solution of the linearized {\em uncoupled} system (Eq.~(\ref{variationaleq})).  Hence the utility of the dispersion function $D(s)$ given by Eq.~(\ref{dispersioneq}) is that it determines the linearized dynamics of the globally coupled system in terms of those of the individual uncoupled systems.  

\section{\label{Discussion}Discussion}

\subsection{\label{Discussion:analytic}Analytic Continuation of $\tilde {\bf M}(s)$}

Consider the $k$th column of $\langle \langle {\bf M}(t)\rangle \rangle_*$, which we denote $[\langle \langle {\bf M}(t)\rangle \rangle_* ]_k$.  According to our definition of ${\bf M}_i$ given by Eq.~(\ref{variationaleq}), we can interpret $[\langle \langle {\bf M}(t)\rangle \rangle_* ]_k$ as follows.  Assume that for each of the uncoupled systems $i$ in Eq.~(\ref{ithuncoupled}), we have a cloud of an infinite number of initial conditions sprinkled randomly according to the natural measure on the uncoupled attractor.  Then, at $t=0$, we apply an equal infinitesimal displacement $\delta_k$ in the direction $k$ to each orbit in the cloud.  That is, we replace ${\bf x}_i(0)$ by ${\bf x}_i(0)+\delta _k{\bf a}_k$, where ${\bf a}_k$ is a unit vector in ${\bf x}$-space in the direction $k$.  Since the particle cloud is displaced from the attractor, it relaxes back to the attractor as time evolves.  The quantity $[\langle \langle {\bf M}\rangle \rangle_* ]_k\delta _k$ gives the time evolution of the $i$-averaged perturbation of the centroid of  the cloud as it evolves back to the attractor and redistributes itself on the attractor.  

We now argue that $\langle \langle {\bf M}\rangle \rangle_* $ decays to zero exponentially with increasing time.  We consider the general case where the support of the smooth density $\rho ({\bf \Omega })$ contains open regions of ${\bf \Omega }$ for which the dynamical system (\ref{ithuncoupled}) has attracting periodic orbits as well as a positive measure of ${\bf \Omega }$ on which Eq.~(\ref{ithuncoupled}) has chaotic orbits.  Numerical experiments on chaotic attractors (including structurally unstable attractors) generally show that they are strongly mixing; i.e., a cloud of many particles rapidly arranges itself on the attractor according to the natural measure.  Thus, for each ${\bf \Omega}_i$ giving a chaotic attractor, it is reasonable to assume that the average of ${\bf M}_i$ over initial conditions ${\bf x}_i(0)$, denoted $\langle {\bf M}_i\rangle_* $, decays exponentially.  For a periodic attractor, however, $\langle {\bf M}_i\rangle_* $ does not decay: a distribution of orbits along a limit cycle comes to the same distribution after one period, and this repeats forever.  Thus, if the distribution on the limit cycle was noninvariant, it remains noninvariant and oscillates forever at the period of the periodic orbit.  On the other hand, periodic orbits exist in open regions of ${\bf \Omega }$, and, when we average over ${\bf \Omega }$, there is the possibility that with increasing time cancellation causing decay occurs via the process of ``phase mixing''\cite{kra73}.  For this case we appeal to an example.  In particular, the explicit computation of $\langle {\bf M}_i\rangle_* $ for a simple model limit cycle ensemble is given in Sec.~\ref{Kuramoto}.  The result is
\[
\langle {\bf M}_i\rangle_* = \frac{1}{2} \left[
\begin{array}{ll} 
\cos \Omega _it & -\sin \Omega _it \\
\sin \Omega _it & \cos \Omega _it
\end{array} \right],
\]
and indeed this oscillates and does not decay to zero.  However, if we average over the oscillator distribution $\rho (\Omega )$ we obtain
\[
\langle \langle \tilde {\bf M}\rangle\rangle_* =\frac{1}{2}\left[
\begin{array}{ll}
c(t) & -s(t) \\
s(t) & c(t)
\end{array}\right],
\]
where $c(t)=\int \rho (\Omega )cos \Omega td\Omega $ and $s(t)=\int \rho (\Omega )\sin \Omega td\Omega $.  For any analytic $\rho (\Omega )$ these integrals decay exponentially with time.  Thus, based on these considerations of chaotic and periodic attractors, we see that for sufficiently smooth $\rho (\bf \Omega )$, there is reason to believe that $\langle \langle {\bf M}\rangle \rangle_* $, the average of ${\bf M}_i$ over ${\bf x}_i(0)$ and over ${\bf \Omega}_i$, decays exponentially to zero with increasing time.  Conjecturing this decay to be exponential, $\big \| \langle \langle {\bf M}(t)\rangle \rangle_* \big \| <\kappa e^{-\xi t}$ for positive constants $\kappa $ and $\xi $, we see that the integral in Eq.~(\ref{mtilde_timeindep}) converges for $\mbox{Re}(s)>-\xi$.  This conjecture is supported by our numerical results in Sec.~\ref{NumericalExperiments}.  Thus, while Eq.~(\ref{mtilde_timeindep}) was derived under the assumption $\mbox{Re}(s)>\Gamma >0$, using analytic continuation, we can regard Eq.~(\ref{mtilde_timeindep}) as valid for $\mbox{Re}(s)>-\xi $.  Note that, for our purposes, it suffices to require only that $\| \langle \langle {\bf M}(t)\rangle \rangle_* \|$ be bounded, rather than that it decay exponentially.  Boundedness corresponds to $\xi =0$, which is enough for us, since, as soon as instability occurs, the relevant root of $D(s)$ has $\mbox{Re}(s)>0$.  

\subsection{\label{Discussion:numerically}Numerically Approximating $\tilde {\bf M}(s)$ by use of Eq.~(\ref{variationaleq})}

We can envision the following numerical method for finding $\langle \langle {\bf M}(t)\rangle \rangle_* $.  First approximate the natural measure of each attractor $i$ by a large {\em finite} number of orbits initially distributed according to the natural measure. For each initial condition, obtain ${\bf x}_i(t)$ from Eq.~(\ref{ithuncoupled}). Then use these solutions in ${\bf DG}$ and solve Eq.~(\ref{variationaleq}). Finally, average the resulting matrix solutions ${\bf M}_i$ over the orbits. While this may look attractive, it can present difficulties in the case of chaotic orbits. In particular, chaos implies that the individual ${\bf M}_i$ diverge exponentially, while the average over the natural invariant measure $\langle {\bf M}_i\rangle_* $ decays.  That is, when we average over the natural invariant measure, the exponential divergence of the individual ${\bf M}_i(t)$ cancel to yield decay. Numerically, however, we average over a large but finite number of orbits.  For early time, one can expect that this will give a good approximation to $\langle {\bf M}_i\rangle_* $. However, as time goes on, the decay of $\langle {\bf M}_i\rangle_* $ implies that the cancellation must become more and more accurate because the individual ${\bf M}_i$ become larger and larger. Thus, eventually, any numerical approximation using a finite number of orbits must diverge. The question is, can one obtain results by this method that are accurate for long enough time to provide a useful basis for approximating $\tilde {\bf M}(s)$. We expand and illustrate this issue in greater detail in Sec.~\ref{NumericalExperiments}.

A variant of the above numerical technique is to obtain $\langle \langle {\bf M}\rangle \rangle_* $ by working directly with the uncoupled nonlinear equations (\ref{ithuncoupled}).  We use a large finite number of initial conditions chosen randomly with respect to $\rho ({\bf \Omega })$ and the natural invariant measure.  These initial conditions are then all displaced, ${\bf x}_i(0)\rightarrow {\bf x}_i(0)+{\bf \Delta } _k$, where ${\bf \Delta}_k=\Delta _k{\bf a}_k$ and $\Delta _k$ is small.  Denoting the solutions of Eq.~(\ref{ithuncoupled}) with these displaced initial conditions ${\bf x}_i'(t, {\bf \Delta }_k)$, we then approximate the quantity $(\langle \langle {\bf x}\rangle \rangle_* - \langle \langle {\bf x}'(t,{\bf \Delta }_k)\rangle \rangle)$, which represents the relaxation of the measure's centroid after a displacement ${\bf \Delta } _k$.  In the limit $\Delta _k\rightarrow 0$, we have that the $k$th column of $\langle \langle {\bf M}(t)\rangle \rangle_* $ is 
\begin{equation}
[\langle \langle {\bf M}(t)\rangle \rangle_* ]_k=\Delta _k^{-1}(\langle \langle {\bf x}\rangle \rangle_* - \langle \langle {\bf x}'\rangle \rangle),
\label{kthcolofm}
\end{equation}
and Eq.~(\ref{kthcolofm}) is used, with small $\Delta_k$, as an approximation. Again, practical numerical issues exist for this technique.  In particular, $\Delta _k$ must be small for linearity to be approximated, but this makes the cancellation of $\langle \langle {\bf x}\rangle \rangle_* - \langle \langle {\bf x}'\rangle \rangle$ stronger, which, in turn, necessitates using many initial conditions.  Also, as time increases, $\langle \langle {\bf M}(t)\rangle \rangle_* $ decreases, and fluctuations of $\langle \langle {\bf x}\rangle \rangle_* - \langle \langle {\bf x}'\rangle \rangle$ due to the finite number of initial conditions can overwhelm the computation of the coherent relaxation to the attractor (see Sec.~\ref{NumericalExperiments}).

\subsection{\label{Discussion:Mtilde}Numerical approximation of $\tilde {\bf M}(s)$ as the Response to $\exp (st) \openone$}

Since $\langle \langle {\bf M}\rangle \rangle_* $ is the response to an impulse drive, its Laplace transform multiplied by $e^{st}$, $\tilde {\bf M}(s)e^{st}$, is the response to a drive $e^{st} \openone$.  We now show this more formally.  First we note that Eq.~(\ref{variationaleq}) with the initial condition ${\bf M}_i = \openone$ at $t=0$ can be written in the impulse response form,
\[
d{\bf M}_i/dt={\bf DG \cdot M}_i+\delta (t) \openone,
\]
where $\delta (t)$ is a delta function, and ${\bf M}_i$ satisfies the initial condition ${\bf M}_i={\bf 0}$ at $t=-\infty$.  Shifting time by $t_0$, we have 
\begin{equation}
\frac{d}{dt}{\bf M}_i(t-t_0,{\bf x}_i(t_0))={\bf DG \cdot M}_i(t-t_0,{\bf x}_i(t_0))+\delta (t-t_0) \openone,
\end{equation}
where we have explicitly included the dependence of ${\bf M}_i$ on time and on the initial state ${\bf x}_i(t_0)$ of the unperturbed orbit ${\bf x}_i(t)$.  Multiplying by $e^{st_{0}}dt_0$ and integrating over all $t_0$ we obtain
\begin{equation}
d\hat{\bf M}_i/dt={\bf DG} \cdot \tilde{\bf M}_i+e^{st} \openone,
\end{equation}
where
\begin{equation}
\hat{\bf M}_i=\int ^t_{-\infty}e^{st_{0}}{\bf M}_i(t-t_0,{\bf x}_i(t_0))dt_0,
\end{equation}
which converges at the lower limit provided $\mbox{Re}(s)$ is sufficiently large.  Introducing $T=t-t_0$, averaging, and, as before, interchanging the order of the average and the integral, we have that the response to $e^{st} \openone$ is, as claimed, $\langle \langle \hat{\bf M}\rangle \rangle_* =e^{st}\tilde {\bf M}(s)$.  This suggests the following numerical technique for finding $\tilde {\bf M}(s)$.  Solve
\begin{equation}
\frac{d \tilde{\bf x}_i^{(c,s)}}{dt}
={\bf G}(\tilde{\bf x}_i^{(c,s)},{\bf \Omega }_i)+{\bf \Delta} _k\bigg\{
\begin{array}{l}
e^{\sigma t}\cos \omega t \\
e^{\sigma t}\sin \omega t,
\end{array}
\label{freqresponse}
\end{equation}
where $s=\sigma -i\omega$, ${\bf \Delta}_k=\Delta _k{\bf a}_k$, and $\Delta _k$ is real. For large $t$, but $\Delta_k e^{\sigma t}$ still small throughout the interval $(0,t)$, we can regard the average response as approximately linear.  Thus the $k$th column of $\tilde {\bf M}(s)$ is 
\begin{equation}
[\tilde {\bf M}(s)]_k\simeq \Delta _k^{-1}e^{-st}[\langle \langle {\bf x}\rangle \rangle_* - \langle \langle \tilde {\bf x}\rangle \rangle],
\label{afterfreqresponse}
\end{equation}
where $\tilde {\bf x}_i=\tilde {\bf x}_i^{(c)}-i\tilde {\bf x}_i^{(s)}$.  Numerically $\langle \langle \tilde {\bf x}\rangle \rangle $ can be approximated using a large finite number of orbits.  In Ref.~\cite{sak00}, a technique equivalent to this with $s$ taken to be imaginary $(s=-i\omega )$ is used to obtain the marginal stability condition (see also Ref.~\cite{top01}).  In Sec.~\ref{NumericalExperiments} we compare the numerical efficacy of this technique and of the techniques discussed in Sec.~\ref{Discussion:numerically} \cite{20}.  The reasoning in Ref.~\cite{sak00} is heuristic, and, adapted to our setting \cite{21}, it goes as follows.  Numerically, it is observed that as the coupling is varied, a Hopf bifurcation occurs.  Thus, for conditions just past the bifurcation, the order parameter variation is sinusoidal, $\langle \langle {\bf x} \rangle \rangle_* - \langle \langle \tilde{{\bf x}} \rangle \rangle  \sim e^{-i \omega t}$. Using this as the drive in the nonlinear equation, as in Eq.~(\ref{freqresponse}), and computing $\tilde{\bf M}(-i\omega)$ as above, self-consistency then yields $\{\openone+\tilde {\bf M}(-i\omega)\cdot {\bf K}\}{\bf \Delta}={\bf 0}$. For the case of a coupling matrix with one nonzero element located on the diagonal [i.e., $K _{11} =k$ and $K_{ij}=0$ if $(i,j)\neq (1,1)]$, the consistency condition then gives $1+\tilde M_{11}(-i\omega)k=0$.  Setting the real and imaginary parts of this equation equal to zero determines the value of the frequency at instability onset and the critical value of the coupling constant $k$ \cite{newref}.  

\subsection{\label{Discussion:distfn}The Distribution Function Approach}

Much previous work has treated the Kuramoto problem and its various generalizations using a kinetic equation approach\cite{str92,erm91,ace00,yeu99,koz00,red98,fra01,str00,wal69,buc88}. We have also obtained our main result, Eq.~(\ref{dispersioneq}) for $D(s)$, by this more traditional method. We briefly outline the procedure below.

Let $F({\bf x},{\bf \Omega },t)$ be the distribution function (actually a generalized function) such that $F({\bf x}, {\bf \Omega },t)d{\bf x}d{\bf \Omega }$ is the fraction of oscillators at time $t$ whose state and parameter vectors lie in the infinitesimal volume $d{\bf x}d{\bf \Omega }$ centered at $({\bf x},{\bf \Omega })$.  Note that $\int Fd{\bf x}$ is time independent, since it is equal to the distribution function $\rho ({\bf \Omega })$ of the oscillator parameter vector.  The time evolution of $F$ is simply obtained from the conservation of probability following the system evolution,
\begin{equation}
\frac{\partial F}{\partial t}+
\frac{\partial}{\partial {\bf x}}\cdot [({\bf G}({\bf x},{\bf \Omega })+{\bf K}\cdot
(\langle \langle {\bf x}\rangle \rangle _*-\langle \langle {\bf x}\rangle
\rangle ))F]=0,
\label{distfunappr}
\end{equation}
where
\begin{equation}
\langle \langle {\bf x}\rangle \rangle =\int \int d{\bf x}d{\bf \Omega }{\bf x}F,
\end{equation}
%
\begin{equation}
\langle \langle {\bf x}\rangle \rangle _*=\int \int d{\bf x}d{\bf \Omega }{\bf x}F_0,
\end{equation}
and $F_0=F_0({\bf x},{\bf \Omega })=f({\bf x},{\bf \Omega })\rho ({\bf \Omega })$, in which $f({\bf x},{\bf \Omega })$ is the density corresponding  to the natural invariant measure of the uncoupled attractor whose parameter vector is ${\bf \Omega }$.  Thus $f({\bf x},{\bf \Omega })$, which is a generalized function, formally satisfies 
\begin{equation}
\frac{\partial}{\partial {\bf x}} \cdot [{\bf G}({\bf x},{\bf \Omega })f({\bf x},{\bf \Omega })]=0.
\end{equation}
Hence, $F=F_0$ is a time-independent solution of Eq.~(\ref{distfunappr}) (the ``incoherent solution'').  We examine the stability of the incoherent solution by linearly perturbing $F$, $F=F_0+\delta F$, to obtain
\begin{equation}
\frac{\partial \delta F}{\partial t}+\frac{\partial}{\partial {\bf x}}\cdot [{\bf G}({\bf x},{\bf \Omega })\delta
F-{\bf K}\langle
\langle \delta {\bf x}\rangle \rangle F_0]=0
\label{distfunappr_2}
\end{equation}
%
\begin{equation}
\langle \langle \delta {\bf x}\rangle \rangle =\int \int d{\bf x}d{\bf \Omega }{\bf x}\delta F.
\label{avgdxasints}
\end{equation}
We can then introduce the Laplace transform, solve the transformed version of Eq.~(\ref{distfunappr_2}), and substitute into Eq.~(\ref{avgdxasints}) to obtain the same dispersion function $D(s)$ as in Sec.~\ref{StabilityAnalysis}. The calculation is somewhat lengthy, involving the formal solution of Eq.~(\ref{distfunappr_2}) by integration along the orbits of the uncoupled system. We will not present the detailed steps here, since the result is the same as that derived in Sec.~\ref{StabilityAnalysis}, where it is obtained in what we believe is a more direct manner.

The distribution function approach outlined above is similar to the marginal stability treatment of Ref.~\cite{top01} for identical globally chaotic maps.  In that case $s\rightarrow -i\omega $, the Frobenius-Perron equation plays the role of Eq.~(\ref{distfunappr}), and the average over parameters is not present.

We note that the computation outlined above is formal in that we treat the distribution functions as if they were ordinary, as opposed to generalized, functions.  In this regard, we note that $f({\bf x},{\bf \Omega })$ is often extremely singular both in its dependence on ${\bf x}$ (because the measure on a chaotic attractor is typically a multifractal) and on ${\bf \Omega }$ (because chaotic attractors are often structurally unstable).  We believe that both these sources of singularity are sufficiently mitigated by the regularizing effect of the averaging process over $({\bf x},{\bf \Omega })$, and that our stability results of Sec.~\ref{StabilityAnalysis} are still valid.  This remains a problem for future study.  We note, however, that for structurally unstable attractors, a smooth distribution of system parameters $\rho ({\bf \Omega })$ is likely to be much less problematic than the case of identical ensemble components\cite{sak00,top01}, $\rho ({\bf \Omega })=\delta ({\bf \Omega }-\bar {\bf \Omega })$.  In the case of identical structurally unstable chaotic components, an arbitrarily small change of $\bar {\bf \Omega }$ can change the character of the base state whose stability is being examined.  In contrast, a small change of a smooth distribution $\rho ({\bf \Omega })$ results in a small change in the weighting of the ensemble members, but would seem not to cause any qualitative change.  We remark that the numerical test cases we study in Sec.~\ref{NumericalExperiments} are all structurally unstable.  Nevertheless, they all agree well with the theory. 

\subsection{\label{Discussion:bifurcations}Bifurcations}

It is natural to ask what happens as a parameter of the system passes from values corresponding to stability to values corresponding to instability.  Noting that the incoherent state represents a time independent solution of Eq.~(\ref{first}), we can seek intuition from standard results on the generic bifurcations of a fixed point of a system of ordinary differential equations (\cite{guc83}; see also \cite{top01}).  There are two linear means by which such a fixed point can become unstable:\ \ (i) a real solution of $D(s)=0$ can pass from negative to positive $s$ values, and (ii) two complex conjugate solutions, $s$ and $s^*$, can cross the imaginary $s$-axis, moving from $\mbox{Re}(s)<0$ to $\mbox{Re}(s)>0$.  

In reference to case (i), we note that the incoherent steady state always exists for our formulation in Sec.~\ref{Formulation}.  In this situation, in the absence of a system symmetry, the generic bifurcation of the system is a transcritical bifurcation (Fig.~\ref{bif_figure}(a)).
\begin{figure}
\caption{\label{bif_figure}Bifurcations. The horizontal line represents the incoherent state. Dashed (solid) lines represent unstable (stable) fixed points, and a system parameter governing the instability increases toward the right.}
\end{figure}
In the presence of symmetry, the existence of a fixed point solution with $\langle \langle {\bf x}\rangle \rangle_* - \langle \langle {\bf x}\rangle \rangle$ nonzero may imply the simultaneous existence of a second fixed point solution with $\langle \langle {\bf x}\rangle \rangle_* - \langle \langle {\bf x}\rangle \rangle$ nonzero, where these solutions map to each other under the symmetry transformation of the system.  In this case the transcritical bifurcation is ruled out, and the generic bifurcation is the pitchfork bifurcation, which can be either subcritical (Fig.~\ref{bif_figure}(b)) or supercritical (Fig.~\ref{bif_figure}(c)).  

In case (ii), where two complex conjugate solutions cross the $\mbox{Im}(s)$ axis, the generic bifurcations are the subcritical and supercritical Hopf bifurcations.  (In this case we note that although the individual oscillators may be behaving chaotically, their average coherent behavior is periodic.)  

In our numerical experiments in Sec.~\ref{NumericalExperiments} we find cases of apparent subcritical and supercritical Hopf bifurcations, as well as a case of what we believe is a subcritical pitchfork bifurcation.  The reason we believe it is a pitchfork rather than a transcritical bifurcation is that our globally coupled system is a collection of coupled Lorenz equations.  Since the Lorenz equations
\begin{equation}
\begin{array}{ccl}
dx^{(1)}/dt & = & \sigma (x^{(2)}-x^{(1)})\\
dx^{(2)}/dt & = & rx^{(1)}-x^{(2)}-x^{(1)}x^{(3)}\\
dx^{(3)}/dt & = & -bx^{(3)}+x^{(1)}x^{(2)}
\end{array}
\label{Lorenz}
\end{equation}
have the symmetry $(x^{(1)},x^{(2)},x^{(3)})\rightarrow (-x^{(1)},-x^{(2)},x^{(3)})$, and since the form of the coupling used in Sec.~\ref{NumericalExperiments} respects this symmetry, the transcritical bifurcation is ruled out.

\subsection{\label{Discussion:generalizations}Generalizations}

One generalization is to consider a general nonlinear form of the coupling such that we replace system (\ref{first}) by 
\begin{equation}
\begin{array}{ccl}
d{\bf x}_i/dt & = & \hat {\bf G}({\bf x}_i,{\bf \Omega }_i,{\bf y}) \\
{\bf y} & = & \langle \langle {\bf x}\rangle \rangle _*-\langle \langle {\bf x}\rangle \rangle
\end{array}
\label{firstgen}
\end{equation}
and the role of the uncoupled system (analogous to Eq.~(\ref{ithuncoupled})) is played by the equation
\begin{equation}
d{\bf x}_i/dt=\tilde {\bf G}({\bf x}_i,{\bf \Omega }_i,{\bf 0}).
\end{equation}
In this more general setting, following the steps of Sec.~\ref{StabilityAnalysis} yields
\begin{equation}
D(s)=\det \{ \openone +\tilde {\bf Q}(s)\},
\label{gendisp}
\end{equation}
where
\[
\tilde {\bf Q}(s)=\int ^\infty _0dTe^{-st}\langle \langle {\bf M}(T){\bf D}_{\bf y}\hat {\bf G}({\bf x},{\bf \Omega
}, {\bf 0})\rangle \rangle_*.
\]

A still more general form of the coupling is
\begin{equation}
d{\bf x}_i/dt=\hat {\hat {\bf G}}({\bf x}_i,{\bf \Omega }_i, \langle \langle {\bf x}\rangle \rangle ).
\label{gentwo}
\end{equation}
For Eqs.~(\ref{firstgen}) and (\ref{first}), a unique incoherent solution $\langle \langle {\bf x}\rangle \rangle _*$ always exists and follows from Eq.~(\ref{xstar}) by solving the nonlinear equations for each ${\bf x}_i(0)$ with ${\bf y}=(\langle \langle {\bf x}\rangle \rangle _*-\langle \langle {\bf x}\rangle \rangle )$ set equal to zero.  In the case of Eq.~(\ref{gentwo}), the existence of a unique incoherent state is not assured.  By definition, $\langle \langle {\bf x}\rangle \rangle $ is time independent in an incoherent state.  Thus replacing $\langle \langle {\bf x}\rangle \rangle $ in Eq.~(\ref{gentwo}) by a constant vector ${\bf u}$, imagine that we solve Eq.~(\ref{gentwo}) for an infinite number of initial conditions distributed for each $i$ on the natural invariant measure of the system, $d{\bf x}_i/dt=\hat {\hat {\bf G}}({\bf x}_i,{\bf \Omega }_i,{\bf u})$, and then use Eq.~(\ref{xstar}) to obtain the average $\langle \langle {\bf x}\rangle \rangle _{\bf u}$.  This average depends on ${\bf u}$, so that $\langle \langle {\bf x}\rangle \rangle _{\bf u}={\bf F}({\bf u})$.  We then define an incoherent solution $\langle \langle {\bf x}\rangle \rangle _*$ for Eq.~(\ref{gentwo}) by setting $\langle \langle {\bf x}\rangle \rangle _{\bf u}={\bf u}=\langle \langle {\bf x}\rangle \rangle_*$, so that $\langle \langle {\bf x}\rangle \rangle _*$ is the solution of the nonlinear equation
\[
\langle \langle {\bf x}\rangle \rangle _*={\bf F}(\langle \langle {\bf x}\rangle \rangle _*).
\]
Generically, such a nonlinear equation may have multiple solutions or no solution.  In this setting, if a stable solution of this equation exists for some paramter $k<k_c$, then the solution of the nonlinear system (\ref{gentwo}) (with appropriate initial conditions) will approach it for large $t$.  If now, as $k$ approaches $k_c$ from below, a real eigenvalue approaches zero, then $k=k_c$ generically corresponds to a saddle-node bifurcation.  That is, an unstable incoherent solution merges with the stable incoherent solution, and, for $k>k_c$, neither exist.  In this case, loss of stability by the Hopf bifurcation is, of course, still generic, and the incoherent solution continues to exist before and after the Hopf bifurcation.  $D(s)$ for Eq.~(\ref{gentwo}) is given by Eq.~(\ref{gendisp}) with ${\bf D}_{\bf y}\hat {\bf G}$ replaced by $-{\bf D}_{\langle \langle {\bf x}\rangle \rangle }\hat {\hat {\bf G}}$ evaluated at the incoherent state $(\langle \langle {\bf x}\rangle \rangle =\langle \langle {\bf x}\rangle \rangle_*)$ whose stability is being investigated.

Another interesting case is when the coupling is delayed by some linear deterministic process.  That is, the ith oscillator does not sense $\langle \langle {\bf x}\rangle \rangle $ immediately, but rather responds to the time
history of $\langle \langle {\bf x}\rangle \rangle $.  Thus, using Eq.~(\ref{firstgen}) as an example, the coupling term ${\bf y}$ is replaced by a convolution,
\[
{\bf y}(t)=\int ^t_{-\infty}dt'{\bf \Lambda}(t-t')\cdot (\langle \langle {\bf x}\rangle \rangle _*-\langle \langle
{\bf x}\rangle \rangle _{t'}).
\]
In this case a simple analysis shows that Eq.~(\ref{gendisp}) is replaced by
\[
D({\bf s})=\det \{ \openone +\tilde {\bf Q}(s)\cdot {\bf \Lambda } (s) \}
\]
where
\[
\tilde {\bf \Lambda }(s)=\int ^\infty _0dte^{-st}{\bf \Lambda}(t').
\]
The simplest form of this would be a discrete delay 
\[
{\bf \Lambda }(t)={\bf K}\delta (t-\eta ),
\]
in which case $\tilde {\bf \Lambda }(s)=\openone e^{-\eta s}$.  (The case of time delayed interaction has been studied for coupled limit cycle oscillators in Refs.~\cite{yeu99,koz00,red98}.)

In addition to these generalizations, others are also of interest.  For example, the inclusion of noise and coupling ``inertia'' is studied in the limit cycle case in Ref.~\cite{ace00}.

\section{\label{Kuramoto}The Kuramoto Problem}

As an example, we now consider a case that reduces to the well-studied Kuramoto problem.  We consider the ensemble members to be two dimensional,
${\bf x}_i=(x_i(t), y_i(t))^T$,
and characterized by a scalar parameter $\Omega_i$.  For the coupling matrix ${\bf K}$ we choose $k\openone$.  Thus Eq.~(\ref{first}) becomes
$dx_i/dt=G^{(x)}(x_i,y_i,\Omega_i)+k(\langle \langle x\rangle \rangle_*
-\langle \langle x\rangle \rangle )$, $dy_i/dt=G^{(y)}(x_i,y_i,
\Omega_i)+k( \langle \langle y \rangle \rangle_* - \langle \langle y \rangle
\rangle )$. We assume that in polar coordinates $(x=r\cos \theta, y=r\sin \theta )$, the uncoupled $(k=0)$ dynamical system is given by
\begin{equation}
d\theta _i/dt=\Omega_i,
\label{anglekura}
\end{equation}
\begin{equation}
dr_i/dt=(r_0-r_i)/\tau,
\label{radiuskura}
\end{equation}
where $\Omega _i\tau \ll 1$.  That is, the attractor is the circle $r_i=r_0$, and it attracts orbits on a time scale $\tau $ that is very short compared to the limit cycle period.  For $\Omega _i\tau \ll 1$ it will suffice to calculate ${\bf M}_i(t)$ for $t\gg \tau $. To do this, as shown in Fig.~\ref{kuramoto_figure}, we consider an initial infinitesimal orbit displacement ${\bf \Delta } _{oi}={\bf a}_xdx_{oi}+{\bf a}_ydy_{oi}$ where ${\bf a}_{x,y}$ are unit vectors.
\begin{figure}
\caption{\label{kuramoto_figure}Illustraion showing how ${\bf M}_i$ is obtained for the Kuramoto example.}
\end{figure}
In a short time this displacement relaxes back to the circle, so that for $(2\pi /\Omega )\gg t\gg \tau $ we have $r=r_0$, $\theta=\theta_{oi}$, ${\bf \Delta }_i(t)\simeq \Delta ^+_{oi}{\bf a}_\theta $, where $\theta _{oi}$ is the initial value $\theta_i(0)$, ${\bf a}_\theta $ is evaluated at $\theta _i(0)$, and $\Delta ^+_{oi}=-\sin \theta _{oi}dx_{oi}+\cos \theta _{oi}dy_{oi}$.  For later time $t\gg \tau $, we have $r=r_0$, $\theta _i(t)=\theta _{oi}+\Omega _it$ and ${\bf \Delta}_i(t)=\Delta ^+_{oi}{\bf a}_\theta $, with ${\bf a}_\theta $ evaluated at $\theta _i(t)$.  In rectangular coordinates this is
\begin{equation}
\Bigg[ \begin{array}{c}
dx_i(t) \\ dy_i(t)
\end{array} \Bigg] =\Bigg[ \begin{array}{ll} \sin (\theta _{oi}+\Omega _it)\sin \theta _{io} & -\sin (\theta_{oi}+\Omega _it)\cos \theta _{io} \nonumber \\[1ex]-\cos (\theta _{oi}+\Omega _it)\sin \theta _{oi} & \cos (\theta _{oi}+\Omega _it)\cos \theta _{oi}\end{array}\Bigg]
\Bigg[ \begin{array}{c} dx_{oi} \\ dy_{oi}\end{array}\Bigg].
\label{bigkura}
\end{equation}
By definition, the above matrix is ${\bf M}_i$ appearing in Sec.~\ref{StabilityAnalysis}.  Averaging Eq.~(\ref{bigkura}) over the invariant measure on the attractor of Eqs.~(\ref{anglekura}) and (\ref{radiuskura}) implies averaging over $\theta _{oi}$.  This yields
\begin{equation}
\langle {\bf M}_i\rangle _\theta =
\frac{1}{2}\Bigg[ \begin{array}{ll}
\cos \Omega _it & -\sin \Omega _it\\ \sin \Omega _it & \cos \Omega _it\end{array}\Bigg].
\end{equation}
Averaging the rotation frequencies $\Omega _i$  over the distribution function $\rho (\Omega )$ and taking the Laplace transform gives $\tilde {\bf M}(s)$, 
\begin{equation}
\tilde {\bf M}(s)=\Bigg[ \begin{array}{ll}
(q_++q_-) & i(q_+-q_-) \\ -i(q_+-q_-)&(q_++q_-)\end{array}\Bigg],
\label{mtildekura}
\end{equation}
where
\begin{equation}
q_\pm (s)=\frac{1}{4}\bigg\langle \frac{1}{s\mp i\Omega }\bigg\rangle _{\Omega }\equiv \frac{1}{4}\int
^{+\infty}_{-\infty } \frac{\rho (\Omega )d\Omega}{s\mp i\Omega },
\label{qpm}
\end{equation}
and, in doing the Laplace transform, we have neglected the contribution to the Laplace integral from the short time
interval $0\leq t\leq 0(\tau )$ (this contribution approaches zero as $\Omega \tau \rightarrow 0$).  Using Eqs.~(\ref{mtildekura}) and (\ref{qpm}) in 
Eq.~(\ref{dispersioneq}) then gives $D(s)=D_+(s)D_-(s)$, where $D_{\pm }(s)$ is the well-known result for the Kuramoto model (e.g.,
\cite{str00}), 
\begin{equation}
D_{\pm}(s)=1+\frac{k}{2}\int ^{+\infty }_{-\infty }
\frac{\rho (\Omega )d\Omega}{s\pm i\Omega } =0\ ,\  Re(s)>0,
\end{equation}
and $D_\pm (s)$ for $\mbox{Re}(s)\leq 0$ is obtained by analytic continuation\cite{str92,str00}.  Note that the property
$D^{\dagger}_{\pm}(s)=D_\mp (s^{\dagger})$, where $\dagger$ denotes complex conjugation, insures that complex roots of $D(s)=D_+(s)D_-(s)=0$ come in conjugate pairs.

\section{\label{NumericalExperiments}Numerical Experiments}

In this section, we illustrate and test our theoretical results using three different ensembles of globally coupled heterogeneous Lorenz oscillators.  The Lorenz equations are given in Eq.~(\ref{Lorenz}).  For our numerical experiments, we set $\sigma=10$ and $b=8/3$, and the ensembles are formed of systems with the parameter $r$ uniformly distributed in an interval $[r_{-},r_{+}]$.  We consider three different cases: an ensemble of periodic oscillators containing a pitchfork bifurcation ($r_-=150$ and $r_+=165$), an apparently chaotic ensemble ($r_-=28$ and $r_+=52$), and an ensemble with mixed chaotic and periodic oscillators ($r_-=167$ and $r_+=202$).

As previously stated, the dispersion function $D(s)$ given by Eq.~(\ref{dispersioneq}) depends only on the solution of the linearized {\it uncoupled} system, and $D(s)$ in turn determines the linear stability of the incoherent state of the {\it globally coupled} system.  To demonstrate this numerically, we consider the simple case in which the coupling matrix ${\bf K}$ has only one nonzero diagonal element, i.e., $K_{11}=k$ and $K_{ij}=0$ for all $(i,j)\neq(1,1)$. For each type of ensemble, a large number $(N\geq10^4)$ of Lorenz equations (Eq.~(\ref{Lorenz})) were initialized with random initial conditions chosen within their respective basins of attraction. The Lorenz equations were integrated using the standard 4th-order fixed-time-step Runge-Kutta method. Each element in the ensemble was independently integrated for a sufficiently long but random time to ensure that the ensemble was essentially on the attractor and was not initiated in a coherent state.  Since the number of oscillators $N$ is large, we choose a simpler form, $\langle\langle x^{(1)} \rangle\rangle = N^{-1}\sum_{i}x^{(1)}_i$, for the order parameter defined in Eq.~(\ref{order}). With $N$ sufficiently large, the average $\langle x^{(1)}_i\rangle$ over the natural measure for a given system $i$ can be absorbed into the summation over $i$.

In the numerical experiments below, we will consider the following time averaged quantity as a measure of the coherence of the order parameter:
\begin{equation}
\label{coherence}
{\bar x}_T=\left[T^{-1}\int^{t+T}_{t}\langle\langle x^{(1)}(t')\rangle\rangle^2 dt'\right]^{1/2},
\end{equation}
where $t$ is sufficiently long so that the ensemble has achieved its time asymptotic behavior, and $T$ is sufficiently long so that ${\bar x}_T$
is essentially independent of $T$. Note that the symmetry of the Lorenz equations under $(x^{(1)}, x^{(2)}, x^{(3)}) \rightarrow (-x^{(1)}, -x^{(2)}, x^{(3)})$ implies that $\langle \langle x^{(1)} \rangle \rangle_* =0$ for $N=\infty$ when the initial conditions are distributed symmetrically in space. Near $k=0$, the ensemble is weakly coupled and the incoherent solution is expected, i.e., ${\bar x}_T \approx0$. (Although $\langle\langle x^{(1)}\rangle\rangle$ is time independent in the infinite $N$ limit for the incoherent state, $\langle\langle x^{(1)} \rangle\rangle$ will exhibit small fluctuations in time for finite $N$.) As the magnitude of $k$ increases, transitions to different coherent states where ${\bar x}_T > 0$ were observed (see Figs.~\ref{fig_periodorder},\ref{fig_chaosorder}, and \ref{fig_mixorder} below). With the three ensembles we have chosen, we have observed super-critical Hopf (the right transition in Fig.~\ref{fig_periodorder} and the left transition in Fig.~\ref{fig_mixorder}), sub-critical Hopf (the left transition in Fig.~\ref{fig_periodorder} and the right transition in Fig.~\ref{fig_mixorder}), and a subcritical pitchfork (Fig.~\ref{fig_chaosorder}) bifurcation from the incoherent state.  

Theoretical predictions for the critical coupling strength $k^*$ for each of these bifurcations can be obtained by estimating ${\tilde M}_{11}(s)$ from the corresponding {\it uncoupled} ensemble.  In particular, we consider the marginal stability condition described in Sec.~\ref{Discussion:Mtilde}.  First, we numerically integrate Eq.~(\ref{freqresponse}) with $\sigma=0$ ($s=-i\omega$) and ${\bf \Delta}=(\Delta,0,0)$.  With $\Delta$ chosen to be sufficiently small, the average response is linear and the $(1,1)$ element of ${\tilde {\bf M}}(-i\omega)$ is
\begin{equation}
\label{miomega}
{\tilde {\bf M}}_{11}(-i\omega) \approx\Delta^{-1}e^{-i\omega t}[\langle\langle x^{(1)} \rangle\rangle_{*} - \langle\langle{\tilde x}^{(1)}\rangle\rangle],
\end{equation}
where $\langle\langle{\tilde x}^{(1)}\rangle\rangle$ is defined in Eq.~(\ref{afterfreqresponse}) and can be obtained numerically from an ensemble of orbits solved from Eq.~(\ref{freqresponse}).  Since we are using a finite number of elements in the ensemble, there will be statistical noise in the ensemble averages calculated on the right hand side of Eq.~(\ref{miomega}); this can be minimized by averaging over time, i.e.,
\begin{equation}
\label{actualm}
{\tilde {\bf M}}_{11}(-i\omega) \approx (T\Delta)^{-1}\int^{T}_{0} e^{-i\omega t}[\langle\langle x^{(1)} \rangle\rangle_{*} - \langle\langle{\tilde x}^{(1)} \rangle\rangle]dt.
\end{equation}
With the coupling matrix ${\bf K}$ being nonzero only in its (1,1) position, Eq.~(\ref{dispersioneq}) yields

\begin{equation}
\label{actuald}
1+{\tilde M}_{11}(-i\omega)k=0.
\end{equation}
The real and imaginary parts of Eq.~(\ref{actuald}) provide two equations that can be used to determine both the critical value of the coupling constant $k^*$ and the frequency $\omega^*$ at the onset of instability. In particular, the imaginary part of the equation, $\mbox{Re}[{\tilde M}_{11}(-i\omega)]=0$, can be solved for $\omega^*$ (note that there may be multiple roots). The real part then yields the corresponding critical coupling $k^*=-[{\tilde M}_{11}(-i \omega^*)]^{-1}$. To determine which of the possibly multiple roots for $\omega^*$ are relevant, we note that as $k$ increases or decreases from zero, a critical value $k^*$ is encountered at which the incoherent state first becomes unstable. Hence we are interested in obtaining the smallest values of $|k^*|$ for both positive and negative $k^*$. (For clarity, we will denote the negative critical value by $-|k^*|$.) Accordingly, the relevant $\omega$ roots are those yielding the largest value of $|{\tilde M}_{11}(-i\omega)|$ for both ${\tilde M}_{11}(-i\omega)>0$ and ${\tilde M}_{11}(-i\omega)<0$. Denoting the corresponding $\omega$ roots by $\omega^*_a$ and $\omega^*_b$, respectively, it is expected that the incoherent state is stable in the range $-|k^*_a|<k<k^*_b$, where
\begin{equation}
\label{kstar}
k^*_{a,b}= - 1/{\tilde M}_{11}(-i\omega^*_{a,b}), \hspace{0.5cm} -|k_a^*| < 0 < k_b^*,
\end{equation}
and that, as $k$ increases through $k^*_b$ (or decreases through $-|k^*_a|$), instability ensues.

Growth rates and frequency shifts from $\omega^*$ can also be simply obtained for $k$ near $k^*$. Setting $k=k^* + \delta k$ and $s = -i(\omega^* + \delta \omega) + \gamma$ in the dispersion relation $1+k{\tilde M}_{11}(s)=0$, and expanding for small $\delta k$, $\delta \omega$ and $\gamma$, we obtain
\begin{equation}
\label{kgrowth}
\gamma = -\frac{\delta k}{(k^*)^2}
\frac{ \partial Im [ {\tilde M}_{11}(-i\omega) ] / \partial \omega }
     {| \partial {\tilde M}_{11} (-i\omega)/ \partial \omega |^2 }
\end{equation}
(and a similar equation for $\delta \omega$), where the expression on the right side of Eq.~(\ref{kgrowth}) is to be evaluated at $\omega = \omega^*$.  Thus, instability implies that $\partial \mbox{Re}\tilde{M}_{11}(-i\omega^*)/\partial\omega <0$ if $k^*,\delta k>0$ and $\partial \mbox{Re}\tilde{M}_{11}(-i\omega^*)/\partial\omega >0$ if $k^*,\delta k<0$.

These growth rates can be estimated numerically by the following procedure. For a given ensemble, we initiate the system in its incoherent state by setting the coupling to zero and integrating for a sufficiently long time. The coupling is then switched on to a value less than $-|k_a|^*$ or larger than $k_b^*$, where the incoherent state is unstable and the order parameter $\langle\langle x^{(1)}\rangle\rangle$ begins to grow exponentially. If the transition is a pitchfork bifurcation, we expect $\langle\langle x^{(1)}\rangle\rangle(t) \sim e^{\gamma t}$ for $\langle\langle x^{(1)}\rangle\rangle$ sufficiently close to the incoherent state. The growth rate $\gamma$ can be obtained by measuring the slope of the graph $\ln \langle\langle x^{(1)}\rangle\rangle(t)$ vs. $t$. If the transition is a Hopf bifurcation, the growth of the order parameter will be modulated by a sinusoidal function. In this case, the envelope of the oscillating order parameter grows exponentially and the growth rate can be extracted by measuring the slope of the logarithm of this envelope function versus time.

\subsection{\label{NumExpts:periodic}Periodic Ensemble}

We first consider an ensemble of Lorenz oscillators with $r_-=150$ and $r_+=165$.  In this range of parameters, the Lorenz equations yield stable periodic orbits.  As $r$ decreases through a critical value $r_c\approx154.7$, the system goes through a pitchfork bifurcation in which a single periodic orbit symmetric under $(x^{(1)}, x^{(2)}, x^{(3)}) \rightarrow (-x^{(1)}, -x^{(2)}, x^{(3)})$ bifurcates into a pair of asymmetric periodic orbits.  The range of dynamics for the (uncoupled, $k=0$) Lorenz equation in this parameter range is illustrated in the bifurcation diagram in Fig.~\ref{fig_periodbif}.
\begin{figure}
\caption{\label{fig_periodbif}Bifurcation diagram for the Lorenz system in the parameter range $r\in[150,165]$ (periodic ensemble).}
\end{figure}
This bifurcation diagram is constructed by plotting the maxima of the function $x^{(3)}(t)$ in the Lorenz equation for $t$ sufficiently large so that any transient is minimized. To further illustrate the pitchfork bifurcation, phase space plots of the limit cycle attractors at $r=165$ and $r=150$ are shown in Figs.~\ref{fig_pitchfork} (a) and (b), respectively.
\begin{figure}
\caption{\label{fig_pitchfork}Periodic orbits from the Lorenz equation with (a) $r=165$ and (b) $r=150$. Black and grey in (b) denote the separate periodic orbits that are present after the pitchfork bifurcation.}
\end{figure}

Figures \ref{fig_periodorder} (a) and (b) show plots of ${\bar x} _{T}$ as a function of the coupling strength $k$.
\begin{figure}
\caption{\label{fig_periodorder}The order parameter as a function of the coupling $k$ for the periodic ensemble. Transitions are observed at (a) $k_b^*=0.95 \pm 0.1$ and (b) $-|k_a^*|=-0.70 \pm 0.04$.}
\end{figure}
$60,000$ oscillators were used. Data are plotted corresponding to the cases in which $k$ decreases (black squares) and $k$ increases (grey circles).  As expected, ${\bar x} _{T}$ is practically zero (to order $O(N^{-1/2})$) for $k$ near 0. As $k$ increases through $k^*_b=0.95 \pm 0.1$, the incoherent state destabilizes and ${\bar x} _{T}$ increases from zero. Similarly, as $k$ decreases through $-|k^*_a|=-0.70 \pm 0.04$, the incoherent state again destabilizes. The transition at $-|k^*_a|$ is hysteretic, while the transition at $k^*_b$ is not. It is also beneficial to examine the time evolution of the instantaneous order parameter $\langle\langle {\bf x}(t)\rangle\rangle$ near the onset of these transitions. This is shown in Figs.~\ref{fig_periodpics} (a) and (b), in which $\langle\langle x^{(2)}(t)\rangle\rangle$ versus $\langle\langle x^{(1)}(t)\rangle\rangle$ is shown before (grey) and just after (black) the transitions at (a) $-|k^*_a|$ and (b) $k^*_b$.
\begin{figure}
\caption{\label{fig_periodpics}Phase portraits showing the transition of the order parameter $\langle\langle {\bf x}(t)\rangle\rangle$ for $k$ slightly past (black) and slightly before (grey) the critical values at (a) $-|k^*_a|$ and (b) $k^*_b$.}
\end{figure}
These transitions are apparently subcritical and supercritical Hopf bifurcations, respectively. The spread in the trajectories is due to the finiteness of $N$; we find that decreasing $N$ increases the spread. In the following, we will investigate the oscillation frequency and growth rate of the order parameter near these transition points.

Using the frequency response method described at the beginning of this section and in Sec.~\ref{Discussion:Mtilde}, we calculated ${\tilde M}_{11}(-i\omega)$ as a function of $\omega$ using an ensemble of uncoupled elements.  We plot both the real (black) and imaginary (grey) parts of ${\tilde M}_{11}(-i\omega)$ in Fig.~\ref{fig_periodm11} (a).
\begin{figure}
\caption{\label{fig_periodm11}
$\mbox{Re}[{\tilde M}_{11}(-i\omega)]$ (black) and $\mbox{Re}[{\tilde M}_{11}(- i\omega)]$ (grey) vs. $\omega$, calculated using (a) the frequency response method ($\Delta=0.05$), and (b) the linear method described in Sec.~\ref{Discussion:numerically}. Both methods yield very similar results overall. Predicted critical coupling values, calculated using the frequency response method and the values at the points indicated in (a), are $-|k_a^*|=-0.72 \pm 0.05$ and $k_b^*=0.93 \pm 0.3$.}
\end{figure}
(For comparison, Fig.~\ref{fig_periodm11} (b) shows the results of the linear displacement method of Sec.~\ref{Discussion:numerically}; see the discussion below.) For these curves, we used a forcing strength $\Delta=0.05$ and $N=20,000$ in our calculations.  As one can see, $\mbox{Re}[{\tilde M}_{11}(-i\omega)]$ crosses zero more than once, and each root corresponds to a possible solution for $\omega^*$. Note that the maxima of $\mbox{Re}[{\tilde M}_{11}(-i\omega)]$ are attained very near, but not necessarily at, these $\omega^*$ roots. The two critical values $-|k^*_a|$ and $k^*_b$ are predicted by Eq.~(\ref{kstar}) with $\omega^*_{a,b}$ corresponding to the largest values of $|\mbox{Re}[{\tilde M}_{11}(-i\omega^*)]|$ for which $\mbox{Re}[{\tilde M}_{11}(-i\omega^*)]=0$.  These values are indicated by the dotted lines in the figure, and yield predictions of 
$-|k^*_a|=-0.72 \pm 0.05$ and $k^*_b=0.93 \pm 0.03$. These predictions agree well with the critical transitions observed in our numerical experiments using the full nonlinear system (see above).
In addition, the predicted frequency at the supercritical bifurcation at $k^*_b$ is $\omega_b^* \approx 21.4$. Fig.~\ref{fig_periodpower} shows the power spectrum of $\langle\langle x^{(1)}(t)\rangle\rangle$ for $k=1.2$, i.e., slightly greater than $k_b^*$; this figure reveals a prominent peak at a frequency of approximately $21.4$, in agreement with the predicted value of $\omega^*_{b}$.
\begin{figure}
\caption{\label{fig_periodpower}
The power spectrum of $\langle\langle x^{(1)}(t)\rangle\rangle$ for $k=1.2$, i.e. slightly larger than $k^*_b$. The largest peak occurs at a frequency of $21.3 \pm 0.1$, in agreement with the predicted value $21.38 \pm 0.15$.}
\end{figure}

Since the elements within this ensemble are not chaotic, the ${\bf M}_i(t)$ do not diverge in time, and we expect the linear displacement method described in Sec.~\ref{Discussion:numerically} for estimating $\langle\langle{\bf M}(t)\rangle\rangle_*$ to work well. The ensemble average $\langle\langle{\bf M}(t)\rangle\rangle_*$ should decay due to ``phase mixing,'' as in the Kuramoto example (see Secs.~\ref{Discussion:analytic} and \ref{Kuramoto}). A graph of $\langle\langle M_{11}(t)\rangle\rangle_*$ for the periodic ensemble is plotted in Fig.~\ref{fig_periodjaco}.
\begin{figure}
\caption{\label{fig_periodjaco}The $(1,1)$ component of $\langle\langle{\bf M}(t)\rangle\rangle_*$ vs. $t$.}
\end{figure}
As one can see, $\langle\langle M_{11}(t)\rangle\rangle_*$ exhibits complex oscillatory behavior as it decays to zero, where small fluctuations presumably due to finite $N$ are evident. To obtain ${\tilde M}_{11}(-i\omega)$, we set $s=-i\omega$ in the Laplace transform of $\langle\langle M_{11}(t)\rangle\rangle_*$. The real (black) and imaginary (grey) parts of ${\tilde M}_{11}(-i\omega)$ (black) obtained by this method are plotted in Fig.~\ref{fig_periodm11} (b). These graphs generally agree with the graphs obtained using the frequency response method (shown in Fig.~\ref{fig_periodm11} (a)) except near the roots of $\mbox{Re}[{\tilde M}_{11}(-i\omega^*)]=0$, where the peaks were not well resolved. Attempts to improve the frequency resolution of the Laplace transform requires a calculation of $\langle\langle M_{11}(t)\rangle\rangle_*$ for longer time. However, fluctuations due to the finite number of ensemble elements prevent the accurate calculation of the decay of $\langle\langle M_{11}(t)\rangle\rangle_*$ to zero for large times. Thus, $N$ must be increased, and practical considerations limit the usefulness of this method (although we note that for this example, the method does yield good values for $\omega_a^*$ and $\omega_b^*$).

Similarly, an accurate measurement of the growth rate of the mean field requires very large ensembles and extremely long transients due to the weak phase mixing, and again we found this calculation to be numerically impractical. Thus, we demonstrate our growth rate calculations only in the computationally more feasible cases considered below, i.e. the chaotic and mixed ensembles.

\subsection{\label{NumExpts:chaotic}A Chaotic Ensemble}

We now consider an ensemble of Lorenz equations with $r_-=28$ and $r_+=52$.  From the bifurcation diagram (see Fig.~\ref{fig_chaosbif}), the ensemble seems to consist of predominantly chaotic systems \cite{allchaos}. 
\begin{figure}
\caption{\label{fig_chaosbif}The bifurcation diagram for the Lorenz system in the parameter range $r\in[28,52]$ (chaotic ensemble).}
\end{figure}
Within this range of parameters, the positive Lyapunov exponent varies between approximately $0.904$ and $1.323$.

Once again, we examined the destabilization of the ensemble's incoherent state by plotting ${\bar x} _{T}$ as a function of $k$.  One can see in Fig.~\ref{fig_chaosorder} that this chaotic ensemble has a hysteretic transition at $-|k^*| = -5.56 \pm 0.01$.
\begin{figure}
\caption{\label{fig_chaosorder}The order parameter as a function of the coupling $k$ for the chaotic ensemble. A subcritical transition is observed at $-|k^*| = -5.56 \pm 0.01$.}
\end{figure}
On the positive side, the incoherent state is stable up to the largest $k$ value tested ($k=7$). Examining the temporal dependence of the instantaneous order parameter $\langle\langle {\bf x}(t)\rangle\rangle$ near the transition at $-|k^*|$, we find that the order parameter jumps to one of two stable fixed points on opposite lobes of the Lorenz attractor (see Fig.~\ref{lorenzlobes}).
\begin{figure}
\caption{\label{lorenzlobes}Phase portrait of the transition of the order parameter (black). The central black oval is before the transition; afterwards, the order parameter shifts to one of the two lateral black dots. A single uncoupled Lorenz attractor for $r=52$ is shown in the background (grey) for comparison.}
\end{figure}
As we have discussed previously (Sec.~\ref{Discussion:bifurcations}), this subcritical transition is expected to be a pitchfork bifurcation rather than a transcritical bifurcation due to the intrinsic symmetry of the Lorenz equations.

We calculated ${\tilde M}_{11}(-i\omega)$ as a function of $\omega$ by examining the uncoupled ensemble under periodic perturbation. For this case, we chose the forcing strength $\Delta$ to be $2$ and $N=20,000$. (We varied the value of $\Delta$ by an order of magnitude from 0.5 to 5 and the result does not seem to change significantly; this indicates that the perturbation is sufficiently linear.) Figure \ref{fig_chaosm11} shows the real and the imaginary parts of ${\tilde M}_{11}(-i\omega)$ versus $\omega$ for this case.
\begin{figure}
\caption{\label{fig_chaosm11}$\mbox{Re}[{\tilde M}_{11}(-i\omega)]$ (black) and $\mbox{Re}[{\tilde M}_{11}(- i\omega)]$ (grey) vs. $\omega$. $\Delta=2$. For comparison, $\mbox{Re}[{\tilde M}_{11}(-i\omega)]$ obtained with the linear (solid circles) and the impulse-response (open circles) methods are included.}
\end{figure}
As compared with the previous example, the frequency response curve is simpler.  $\mbox{Re}[{\tilde M}_{11}(-i\omega)]$ crosses zero only at $\omega^*=0$, where $\mbox{Re}[{\tilde M}_{11}(-i\omega)]$ has a prominent peak. Using Eq.~(\ref{kstar}), this gives a critical coupling value of $-|k^*|=-5.57 \pm 0.15$.  This result agrees well with the threshold of instability for the incoherent state observed in the globally coupled ensemble.

We have also attempted to calculate ${\tilde M}_{11}(-i\omega)$ for this (chaotic) ensemble using the other two methods described in Sec.~\ref{Discussion:numerically}. These are: the linear method, in which the linearized equation [Eq.~(\ref{variationaleq})] is solved for $M_{11}(t)$ and the result is averaged, and the impulse-response method, in which the orbits on the attractor are displaced by a small amount in the $x^{(1)}$ direction and the rate of decay back to zero is measured. The results from these methods are included in Fig.~\ref{fig_chaosm11} with filled and open circles, respectively. While all methods agree reasonably well for $\omega > 2.5$, the important narrow peak at $\omega =0$ is missing from the results of both the linear and the impulse-response methods.

This discrepancy can be understood by observing that the peak at $\omega=0$ represents long-time dynamics. In particular, the half-width of this peak has $\Delta \omega \approx 1$, corresponding to a decay time scale of $1/\Delta \omega \approx 1$. In contrast, the spectrum, with this peak deleted, has a half-width of $\Delta \omega \approx 8$, corresponding to a much shorter time scale of approximately $0.1$. The linear and impulse-response methods apparently resolve the short time scale well, but fail to adequately resolve the longer time scale. This is due to the problem that we have discussed in Sec.~\ref{Discussion:numerically}. For the linear method, the individual ${\bf M}_{i}(t)$ grow exponentially in time, and hence the ensemble average $\langle\langle{\bf M}(t)\rangle\rangle_*$ requires a delicate canceling in order to remain valid for large time. Figure \ref{fig_chaosm11t} shows a graph of $\langle\langle M_{11}(t)\rangle\rangle_*$ for the linear method in grey.
\begin{figure}
\caption{\label{fig_chaosm11t}Graph of $\langle\langle M_{11}(t)\rangle\rangle_*$ for the chaotic ensemble using the linear (grey) and the impulse response methods (black).}
\end{figure}
$\langle\langle M_{11}(t)\rangle\rangle_*$ initially decays exponentially, but for $t>0.7$, it begins to grow as the balanced canceling breaks down due to the fact that only a finite number of elements is used in the calculation. Thus, when obtaining the Laplace transform, we only integrated over the reliable range, i.e. $0 \leq t \leq 0.7$. This had the effect of leaving out the slower decay, which is vital for determining the critical coupling strength for the onset of instability in this case. In contrast, when $\langle\langle M_{11}(t)\rangle\rangle_*$ is measured using the impulse response method, it does not ultimately diverge exponentially. However, its exponential decay is masked by fluctuations for $t \geq 0.7$, again due to finite $N$; see the black curve in Fig.~\ref{fig_chaosm11t}. 

We found that the frequency response method is more reliable because the temporal averaging effectively reduces statistical noise. Therefore, we were able to obtain a good estimate of ${\tilde M}_{11}(-i\omega)$ with only a moderate number of oscillators. The cost for these improved statistics is that each calculation is for only one particular value of $\omega$. This is in contrast to the impulse response method, in which the Laplace transform of $\langle\langle M_{11}(t)\rangle\rangle_*$ gives the entire dependence of ${\tilde M}_{11}(-i\omega)$ on $\omega$ at once. Some of the comparative advantages and drawbacks among the three numerical methods in calculating ${\tilde M}_{11}(-i\omega)$ can be clearly seen in this example. 

The growth rate of the incoherent state, when it first becomes unstable, can be estimated from $\partial Im [ {\tilde M}_{11}(-i\omega) ]/ \partial \omega $ at $\omega=\omega^*=0$. According to Eq.~(\ref{kgrowth}), this growth rate ($\gamma$) is a linear function of $\delta k$ for $k$ close to $k^*$. To verify this, we obtained growth rates for various values of $k$ by first initializing the ensemble in the incoherent state, and then fitting a line to the graph of the $ln \langle \langle x^{(1)} \rangle \rangle$ versus time. Since the transition is subcritical, only the initial growth rate is measured. Figure \ref{fig_chaosgrowth} shows the typical behavior for $k$ slightly beyond the critical value.
\begin{figure}
\caption{\label{fig_chaosgrowth}
Graph of $ln \langle\langle x^{(1)}(t)\rangle\rangle$ vs. $t$ showing the destabilization of the incoherent state for $k$ slightly larger than $k^*$. Since the transition is subcritical, only the initial growth rate is estimated as shown.}
\end{figure}
A plot of estimated growth rates versus $\delta k$ is shown in Fig.~\ref{fig_chaosgamma}.
\begin{figure}
\caption{\label{fig_chaosgamma}
$\gamma$ vs. $\delta k$ for the chaotic ensemble near the transition. The slope predicted using the frequency response method is also shown (lines).}
\end{figure}
The predicted slope, calculated from the frequency response method using Eq.~(\ref{freqresponse}), is $-0.29 \pm 0.02$. This agrees well with the measured growth rates.

\subsection{\label{NumExpts:mixed}A Mixed Chaotic Ensemble with Periodic Windows}

For our last example, we consider an ensemble which contains both chaotic and periodic oscillators ($r_-=167$ and $r_+=202$). As one can see from the bifurcation diagram in Fig.~\ref{fig_mixbif}, there is a prominent period-four window near $r=181$.  
\begin{figure}
\caption{\label{fig_mixbif}
The bifurcation diagram for the Lorenz system in the parameter range $r\in[167,202]$ (mixed ensemble).}
\end{figure}
Thus, the chaotic attractors in this ensemble are expected to be structurally unstable. Figure \ref{fig_mixorder} shows a plot of ${\bar x}_T$ as a function of $k$.
\begin{figure}
\caption{\label{fig_mixorder}
The order parameter as a function of the coupling $k$ for the chaotic ensemble. Transitions are observed at $-|k_a^*|=-0.68 \pm 0.03$ and $k_b^*=1.75 \pm 0.05$.}
\end{figure}
The incoherent state becomes unstable as $k$ increases through $k^*_b=1.75 \pm 0.05$ and as $k$ decreases through $-|k^*_a|=-0.68 \pm 0.03$.  For the left (negative) transition, the order parameter $\langle\langle {\bf x}(t)\rangle\rangle$ becomes periodic, and the amplitude of its oscillation gradually increases as $k$ moves beyond its critical value at $-|k_a^*|$. Thus, this transition is a supercritical Hopf bifurcation. The frequency of oscillation of the order parameter near this transition was estimated from the power spectrum of $\langle\langle x^{(1)}(t)\rangle\rangle$ to be $\omega^*_a \approx 12.2$. The transition at $k_b^*$ appears to be a (hysteretic) subcritical Hopf bifurcation. Phase portraits for $k$ on either side of the two transitions are shown in Figs.~\ref{fig_mixpic} (a) and (b).
\begin{figure}
\caption{\label{fig_mixpic}
Phase portraits showing the transition of the order parameter $\langle\langle {\bf x}(t)\rangle\rangle$ for $k$ slightly past (black) and slightly before (grey) the critical values at (a) $-|k^*_a|$ and (b) $k^*_b$.}
\end{figure}

As before, these two transitions can be predicted from ${\tilde M}_{11}(-i\omega)$ calculated from the uncoupled ensemble.  Figure \ref{fig_mixm11} is a graph of the real and the imaginary parts of ${\tilde M}_{11}(-i\omega)$ obtained using the frequency response method ($\Delta=0.7$ and $N=20,000$).
\begin{figure}
\caption{\label{fig_mixm11}
$\mbox{Re}[{\tilde M}_{11}(-i\omega)]$ (black) and $\mbox{Re}[{\tilde M}_{11}(-i\omega)]$ (grey) vs. $\omega$; $\Delta=0.7$ and $N=20,000$. The predicted transition values are $-|k_a^*|=-0.72 \pm 0.09$ and $k_b^*=1.64 \pm 0.08$.}
\end{figure}
As in the periodic ensemble case, the maxima of $\mbox{Re}[{\tilde M}_{11}(-i\omega)]$ occur very near, but not exactly at the $\omega$ roots of $\mbox{Re}[{\tilde M}_{11}(-i\omega)]=0$. Using Eq.~(\ref{kstar}), the values of $\mbox{Re}[{\tilde M}_{11}(-i\omega^*_{a,b})]$ near the two biggest peaks give $-|k^*_a|=-0.72 \pm 0.09$ and $k^*_b=1.64 \pm 0.08$. The predicted transition frequency associated with the supercritical transition at $-|k^*_a|$ is approximately $12.2$. These predictions agree well with the observed quantities obtained using the fully nonlinear, globally coupled ensemble.

We have also compared the actual growth rate obtained from the globally coupled ensemble with its predicted value calculated from ${\tilde M}_{11}(-i\omega)$ using the same procedure described above. Figure \ref{fig_mixgamma} is a graph of $\gamma$ vs. $\delta k$ for the transition at $-|k^*_a|$.
\begin{figure}
\caption{\label{fig_mixgamma}$\gamma$ vs. $\delta k$ for the transition in the mixed ensemble near $-|k^*_a|$. The slope predicted from the frequency response method is also shown (lines).}
\end{figure}
The predicted slope, calculated using the frequency response method using Eq.~(\ref{freqresponse}), is $-0.26 \pm 0.05$; this agrees well with the measured growth rates. 

\section{\label{Further}Conclusion}

We have presented a general formulation for the determination of the stability of the incoherent state of a globally coupled system of continuous time dynamical systems. This formulation gives the dispersion function in terms of a matrix $\tilde{\bf M}(s)$ which specifies the Laplace transform of the time evolution of the centroid of the {\it uncoupled} (${\bf K}=0$) ensemble to an infinitesimal displacement. Thus the stability of the coupled system is determined by properties of the collection of individual uncoupled elements. The formulation is valid for any type of dynamical behavior of the uncoupled elements. Thus ensembles whose members are periodic, chaotic, or a mixture of both can be treated. We discuss the analytic properties of $\tilde{\bf M}(s)$ and its numerical determination. We find that these are connected: analytic continuation of $\tilde{\bf M}(s)$ to the $\mbox{Im}(s)$ axis is necessary for the application of the analysis, but, in the chaotic case (as discussed in Secs.~\ref{Discussion:Mtilde} and \ref{NumericalExperiments}) leads to numerical difficulties in obtaining $\tilde{\bf M}(s)$. We illustrate our theory by application to the Kuramoto problem and by application to three different ensembles of globally coupled Lorenz systems. In particular, our Lorenz ensembles include a case where all of the uncoupled ensemble members are periodic with a pitchfork bifurcation of the uncoupled Lorenz equations within the parameter range of the ensemble, a case where all the ensemble members appear to be chaotic, and a case where the parameter range of the ensemble yields chaos with a window of periodic behavior. These numerical experiments illustrate the validity of our approach, as well as the practical limitations to numerical application.

\begin{acknowledgments}
E.O. acknowledges support from ONR (physics) and NSF (PHYS-0098632); P.S. acknowledges support from NSF (IBN 9727739); E.B. acknowledges support from NIH (K25MH01963).
\end{acknowledgments}


\begin{thebibliography}{}
\bibitem{newref}
Another possible technique that appears to be attractive for ensembles of units whose uncoupled dynamics is chaotic is based on the unstable periodic orbits embedded in the chaotic attractors in conjunction with cycle expansions. This approach is currently under investigation.
\bibitem{allchaos} Narrow windows most likely exist within this parameter range \cite{barreto97}, but they appear be too small to be easily detectable.
\end{thebibliography}
\end{document}